\documentclass[conference]{IEEEtran}
\IEEEoverridecommandlockouts

\usepackage{amsmath,amsfonts,amssymb}
\usepackage{booktabs}
\usepackage{multirow}
\usepackage{graphicx}
\usepackage{algorithm}
\usepackage{algorithmic}
\usepackage{colortbl}
\usepackage{array}
\usepackage{subcaption}
\usepackage{balance}
\usepackage{multirow}
\usepackage[table]{xcolor}
\usepackage{hyperref}

\definecolor{tableheader}{RGB}{46,134,171}
\definecolor{tablerowalt}{RGB}{245,248,250}
\definecolor{bestresult}{RGB}{46,134,171}
\definecolor{lightblue}{RGB}{230,240,255}

\def\BibTeX{{\rm B\kern-.05em{\sc i\kern-.025em b}\kern-.08em
    T\kern-.1667em\lower.7ex\hbox{E}\kern-.125emX}}
\begin{document}

\title{CaST-POI: Candidate-Conditioned Spatiotemporal Ranking for Next POI Recommendation}

\author{\IEEEauthorblockN{1\textsuperscript{st} Zhenyu Yu}
\IEEEauthorblockA{\textit{College of Computer Science and}\\
\textit{Artificial Intelligence}\\   
\textit{Fudan University}\\
Shanghai, China \\
yuzhenyuyxl@foxmail.com}
\and
\IEEEauthorblockN{2\textsuperscript{nd} Chunlei Meng}
\IEEEauthorblockA{\textit{College of Intelligent Robotics}\\
\textit{and Advanced Manufacturing}\\   
\textit{Fudan University}\\
Shanghai, China \\
clmeng23@m.fudan.edu.cn}
\and
\IEEEauthorblockN{3\textsuperscript{rd} Yangchen Zeng}
\IEEEauthorblockA{\textit{School of Cyber Science and}\\
\textit{Engineering}\\   
\textit{Southeast University}\\
Nanjing, China \\
220245765@seu.edu.cn}
\and
\hspace{3.2em}
\IEEEauthorblockN{4\textsuperscript{th} Mohd Yamani Idna Idris}
\IEEEauthorblockA{
\hspace{3.2em}
\textit{Faculty of Computer Science and}\\
\hspace{3.2em}
\textit{Information Technology}\\
\hspace{3.2em}
\textit{Universiti Malaya}\\
\hspace{3.2em}
Kuala Lumpur, Malaysia \\
\hspace{3.2em}
yamani@um.edu.my}
\and
\IEEEauthorblockN{5\textsuperscript{th} Jihong Guan\textsuperscript{*}}
\IEEEauthorblockA{\textit{School of Computer Science and}\\
\textit{Technology}\\
\textit{Tongji University}\\
Shanghai, China \\
jhguan@tongji.edu.cn}
\and
\IEEEauthorblockN{6\textsuperscript{th} Shuigeng Zhou\textsuperscript{*}\thanks{\textsuperscript{*}Corresponding authors.}}
\IEEEauthorblockA{\textit{College of Computer Science and}\\
\textit{Artificial Intelligence}\\
\textit{Fudan University}\\
Shanghai, China \\
sgzhou@fudan.edu.cn}
}

\maketitle

\begin{abstract}
Next Point-of-Interest (POI) recommendation ranks a user's likely next location based on check-in history. Most recent rankers compress the trajectory into a single user vector and score every candidate through the same representation, ignoring that every candidate carries geographic coordinates and that the relevance of a past visit depends on where the candidate is located. Target attention from click-through-rate prediction conditions the user representation on the scored item, but its operator was developed for web items without geometry. This paper proposes \textbf{CaST-POI}, a candidate-conditioned POI ranker that keeps a standard target-attention reader and adds two bucketised biases to the attention logits: one for the recency of each past visit and another for its distance to the candidate being scored. Different candidates therefore read the same trajectory with different attention weights, grounded in real geographic distance rather than inferred from item embeddings. Comparing against seven sequential recommenders trained with the same split on three datasets NYC, TKY and CA under per-user leave-one-out with full-vocabulary ranking, CaST-POI improves MRR over the strongest baseline by $5.1\%$, $7.6\%$ and $14.5\%$. Holm-corrected paired tests confirm the gains on TKY and CA. Ablation study locates the gain in the explicit revisit gate and the candidate-relative spatial bias, whereas the candidate-independent temporal bias has no measurable effect on all datasets. Code is available at https://github.com/YuZhenyuLindy/CaST-POI.
\end{abstract}

\begin{IEEEkeywords}
Sequential Recommendation; Point-of-Interest Recommendation; Candidate-Conditioned Ranking; Target Attention; Spatiotemporal Modeling.
\end{IEEEkeywords}

\begin{figure}[t]
\centering
  \includegraphics[width=0.99\linewidth]{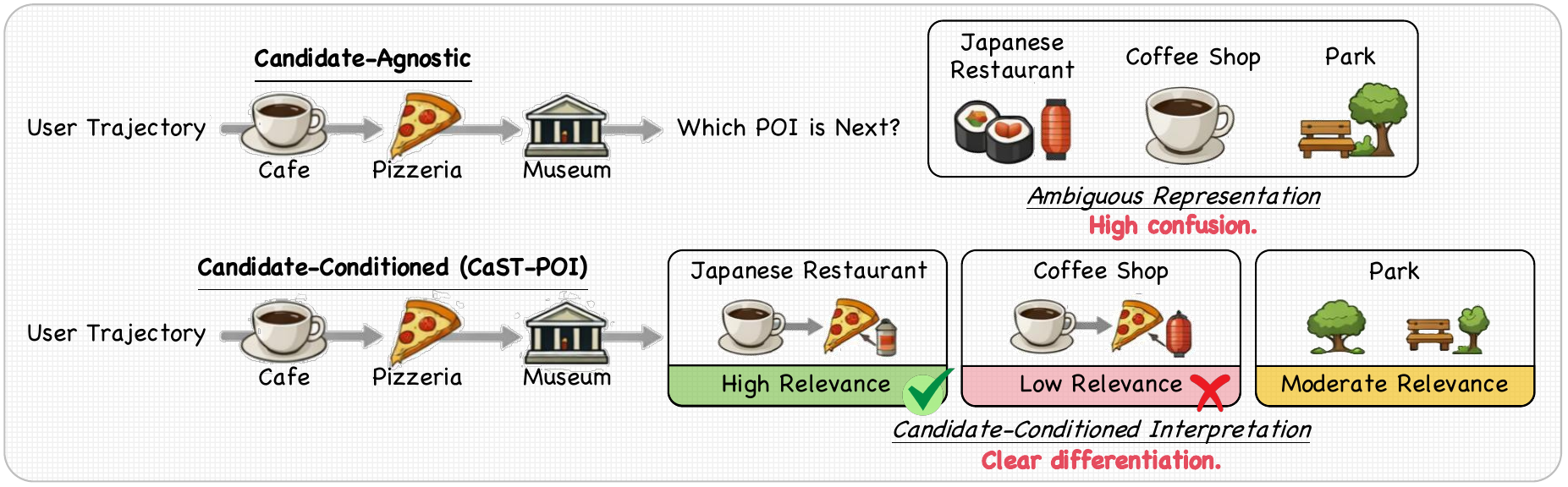}
  \caption{Candidate-agnostic vs.\ candidate-conditioned ranking. Traditional methods compute a single user representation with which to score all candidates. CaST-POI instead uses each candidate as a query to read the same user history differently, producing candidate-specific representations.}
  \label{fig:motivation}
\end{figure}

\section{Introduction}
Suppose a user frequently visits coffee shops near home in the morning, a caf\'e near the office at lunch, and a few restaurants in a third neighborhood on weekday evenings. A ranker scoring a specific Japanese restaurant as the next POI has to decide which parts of this history are relevant, and the answer depends on where the Japanese restaurant actually is. If it sits a few blocks from the office and the time is evening, the user's weekday evening dining in that area is the informative signal, and the morning coffee visits are beside the point. If it is close to home, the relevant subset shifts toward weekend patterns. If it is in a part of the city the user has never visited, the lack of nearby prior visits is itself informative. The same trajectory, read against different candidates, calls for different interpretations. A model that compresses the whole history into one fixed vector cannot produce all those readings at once, so it must commit to a single averaged reading that fits no candidate particularly well.

Candidate-conditioned user representation is not a new idea. In click-through-rate (CTR) prediction, target attention~\cite{zhou2018deep,zhou2019deep,pi2020search,yu2026ads} makes the user representation depend on the candidate being scored and consistently beats candidate-independent pooling; yet mainstream POI rankers, from RNN-based mobility models through graph-enhanced sequential rankers~\cite{yang2022getnext,huang2024learning,yu2026ghost,yu2026mars} to recent LLM-based approaches~\cite{li2024large,zhong2025comapoi}, still encode a trajectory once and reuse it for every candidate. Importing that operator would not by itself close the gap, because POI is not the same setting as CTR. Its candidate pool is large, often tens of thousands of locations inside one city or region, so a single compressed history has to separate many competing options at once, many of them close together and sharing categories. The sharper difference, though, is geometry. Each candidate POI and each historical visit carry explicit coordinates, and the distance between them is a direct predictive signal rather than something the model must infer from item embeddings: a candidate 200m from a frequent past location and one 20km away engage very different subsets of a user's history, even when they belong to the same category. Target attention was developed for products in a catalogue, where no canonical distance relates a candidate to a past click, so its operator never had to represent statements of the form \textit{this candidate sits near that past visit}, and has no channel through which this geometry can enter the attention weights. A POI ranker that borrows the operator alone inherits this blind spot.

Addressing this gap does not require a new attention operator, only direct access to the candidate-relative geometry. In this paper we propose \textbf{CaST-POI} --- a \textbf{\underline{Ca}}ndidate-conditioned \textbf{\underline{S}}patio\textbf{\underline{T}}emporal \textbf{\underline{POI}} ranker. It keeps target attention as its sequence reader and adds two bucketised biases to the attention logits. A temporal bias encodes how recent each historical visit is, and a candidate-relative spatial bias encodes how close each historical visit lies to the candidate currently being scored. With these two biases, different candidates attend to the same history with different weights, and those weights are grounded in real geographic distance rather than inferred from item embeddings alone. The reader supplies a refinement on top of a candidate-agnostic dot-product base and an explicit revisit gate, so that its contribution is measured as a marginal gain over a strong scorer rather than against a weak one. By deliberately keeping the architectural change minimal, the gains can be attributed to exposing the geometry rather than to a richer model.
Our \textbf{contributions} are as follows:
\begin{itemize}
    \item \textbf{A candidate-conditioned view of next-POI ranking.}
    Current POI rankers remain candidate-agnostic despite the task being inherently candidate-dependent. Importing CTR-style target attention alone does not close this gap, because POI candidates carry explicit geometric structure that the attention operator cannot express.

    \item \textbf{Spatiotemporally grounded target attention in a single forward pass.}
    We add two bucketised biases to the attention logits, one for the recency of each past visit and one for its distance to the candidate being scored. They cost negligible parameters and share one history encoding across all candidates, so the whole vocabulary is scored without the multi-stage retrieval or LLM-based reasoning that previous candidate-aware POI methods rely on.

    \item \textbf{An extensive evaluation that supports the claim.}
    Every number comes from our own runs on one leave-one-out split with full-vocabulary ranking, three seeds, and paired significance tests with multiple-comparison correction. We report where the method does \emph{not} win, and include a zero-parameter revisit heuristic that proves a stronger reference than several published neural rankers.
\end{itemize}

\section{Related Work}

\subsection{Candidate-Agnostic Sequential Rankers}

Next POI recommendation has evolved from collaborative filtering to deep sequential models. Recurrent approaches such as GRU4Rec~\cite{hidasi2016session} treated user behaviour as a sequence, while STRNN~\cite{liu2016predicting} and DeepMove~\cite{feng2018deepmove} added spatial and temporal signals to them. Self-attention followed with SASRec~\cite{kang2018self} and BERT4Rec~\cite{sun2019bert4rec}, and POI-specific models such as GETNext~\cite{yang2022getnext} and MTNet~\cite{huang2024learning} extended the line to spatial structure and mobility dynamics. In these architectures the trajectory is encoded into a single fixed representation \emph{before} any candidate is considered, forcing one reading of the history to score each candidate at once, however the candidates differ much.

Alongside this line, general session- and sequence-based recommenders are routinely applied to check-in data and form the baselines of this paper:  NARM~\cite{li2017neural} with attention pooling over a recurrent encoder, STAMP~\cite{liu2018stamp} with a short-term attention/memory priority module, Caser~\cite{tang2018personalized} with horizontal and vertical convolutions over the recent sequence, SR-GNN~\cite{wu2019session} with a gated GNN over session graphs, and CORE~\cite{hou2022core} with session encodings and item embeddings constrained to one representation space. They make no POI-specific assumption, which is why they isolate what a strong candidate-agnostic encoder achieves before spatial structure is added. Refinements to this pipeline, whether in the attention weights or in negative sampling, help a fixed representation fit the data better, but its shape still does not change across candidates.

\subsection{Spatiotemporal and Structural Modelling}

Temporal information in POI models is typically encoded as absolute features such as hour-of-day embeddings or as relative features such as inter-visit intervals, and spatial information through distances, graphs, or learned spatial representations, as in STAN~\cite{luo2021stan} and Graph-Flashback~\cite{rao2022graph}. Recent work enriches this line further, aligning collaborative, transitional, and geographical hypergraph views by contrastive learning~\cite{lai2024disentangled} or exposing which history positions drive each prediction through a variational information bottleneck~\cite{yang2024self}. These approaches enrich the spatiotemporal representation of a trajectory, yet still model relevance independently of the candidate under evaluation. Features are encoded at the trajectory level rather than as candidate-relative quantities, so the same signal carries equal weight against every candidate, which conflicts with the candidate-dependent nature of mobility decisions.

\subsection{Candidate-Conditioned Representation}

Target attention uses the candidate item as a query to attend over past behaviour, so the user representation depends on what is being scored. The idea originated in click-through-rate prediction. DIN~\cite{zhou2018deep} obtains a candidate-specific user representation through an activation-unit attention over behaviour sequences, DIEN~\cite{zhou2019deep} precedes it with a GRU-based interest evolution layer, BST~\cite{chen2019behavior} replaces the activation unit with Transformer self-attention while keeping the candidate-as-query design, and SIM~\cite{pi2020search} retrieves a relevant subset first so that candidate-conditioned modelling remains feasible on very long sequences. Later work extends the operator to pattern-level dependencies and, through consistency-preserved retrieval, to lifelong behaviour sequences~\cite{si2024twin}. Across these designs, candidate-dependent user representations outperform a single pre-computed embedding.

These designs were developed for CTR, where candidates share no geometric relationship with past behaviour, so the attention weights never had to express that a candidate lies near a past visit. POI exposes that geometry, but a ranker inheriting only the CTR operator must learn it indirectly through item embeddings or co-occurrence. Within POI, candidate-awareness appears in partial forms. ROTAN~\cite{feng2024rotan} conditions temporal attention on the target timestamp through a rotation operator, exposing one candidate-relative quantity while leaving target location implicit. An LLM-based line includes SeCor~\cite{wang2024secor}, which aligns semantic and collaborative representations by prompting, GNPR-SID~\cite{wang2025generative}, which formulates next-POI as generative modelling over residual-quantised semantic IDs, and CoMaPOI~\cite{zhong2025comapoi}, which constrains the candidate space by multi-agent reasoning and places candidate-awareness in the pipeline rather than the representation, producing no reusable candidate-conditioned encoding. CaST-POI instead places candidate-awareness inside representation learning and exposes candidate-relative temporal and spatial structure through the attention logits.

\section{Problem Formulation}

Let $\mathcal{U}$ be the set of users and $\mathcal{P}$ the set of POIs where each $p\in\mathcal{P}$ has an identifier and a geographic location $\ell_p=(\text{lat},\text{lon})$. A check-in is a triple $(u,p,t)$ of user, visited POI and timestamp. For each user $u$, the time-ordered history is
\begin{equation}
\mathcal{H}_u = \big\langle (p_1,t_1),(p_2,t_2),\ldots,(p_L,t_L) \big\rangle ,
\qquad t_1 \le \cdots \le t_L ,
\end{equation}
so that $(p_L,t_L)$ is the most recent visit. Histories are truncated to the most recent $L$ visits and left-padded, so $L$ denotes this fixed reader window, and we write $\ell_i$ for the location of $p_i$ when it must be referenced separately from the POI identifier.

Given a candidate set $\mathcal{C}_u \subseteq \mathcal{P}$, next-POI ranking assigns each candidate a score reflecting its likelihood of being the user's next visit, through a learned scoring function
\begin{equation}
s(u,c) = f(\mathcal{H}_u, c), \qquad c \in \mathcal{C}_u .
\end{equation}
In candidate-agnostic rankers, $f$ pools $\mathcal{H}_u$ into a fixed user representation and consults $c$ only at the scoring step. Here $f$ conditions on $c$ while reading $\mathcal{H}_u$, so the same history can be read differently for different candidates, which is the formal property the method exploits. Throughout our experiments $\mathcal{C}_u=\mathcal{P}$, that is, every POI is scored; Section~\ref{sec:candidate_pool_size} examines what happens when $\mathcal{C}_u$ is a sampled subset instead.

\begin{figure*}
    \centering
    \includegraphics[width=0.80\linewidth]{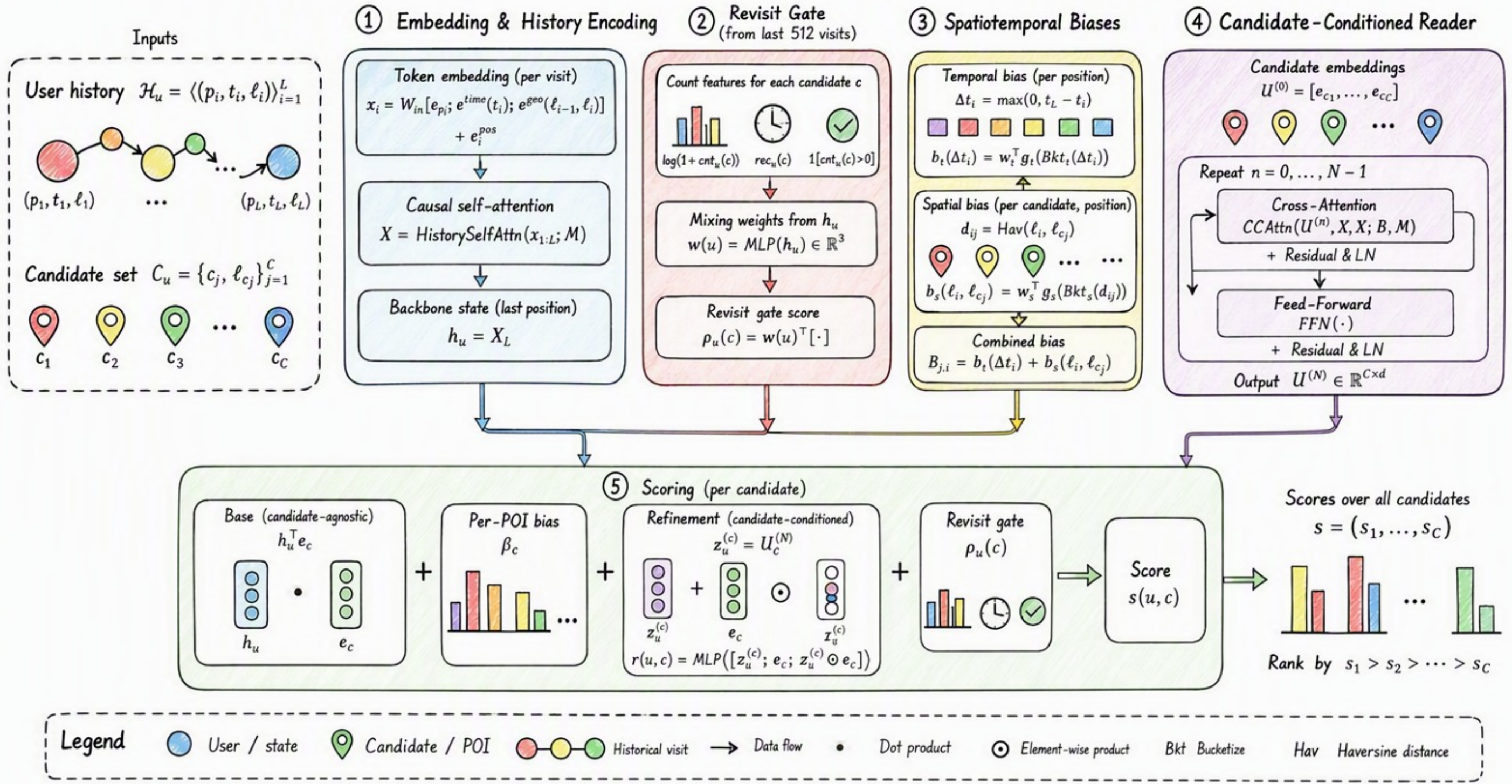}
    \caption{CaST-POI framework and Workflow. The trajectory is encoded once (1) and read by candidate queries through cross-attention (4), whose logits carry a candidate-relative spatial bias and a temporal bias (3). The final score of Eq.~\eqref{eq:score} sums a candidate-agnostic base, a per-POI bias, the candidate-conditioned refinement, and the revisit gate (2), and all candidates are ranked in one forward pass (5).}
    \label{fig:overview}
\end{figure*}

\section{The CaST-POI Method}
\label{sec:method}

\subsection{Overview}
As shown in Fig.~\ref{fig:overview}, CaST-POI scores each candidate $c_j\in\mathcal{C}$ through a representation of the user history $\mathcal{H}$ that depends on which candidate is being scored. Its reader is a standard target-attention block with candidate embeddings as queries over the same trajectory, and the rest of the network is a standard Transformer stack; the novelty is not the operator but the bucketised biases that inject candidate-relative geometry into its attention logits. The reader is also not the whole scorer: next-POI data is revisit-dominated, and a model with no explicit channel for how often and how recently a user has visited a candidate leaves that signal to be rediscovered by the embeddings. CaST-POI therefore sums four terms: a dot product between a causal self-attention state and the candidate embedding, a per-POI bias, the candidate-conditioned refinement that is this paper's contribution, and an explicit revisit gate. Section~\ref{sec:scoring} defines the sum, and Section~\ref{sec:ablation} measures each term separately.

\subsection{Embedding and History Encoding}
\label{sec:41}
Each check-in is turned into a token by concatenating three feature groups and projecting the result back to the model width $d$, after which a learned positional embedding is added,
\begin{equation}
\mathbf{x}_i = \mathbf{W}_{\text{in}}\big[\mathbf{e}_{p_i};\,\mathbf{e}^{\text{time}}(t_i);\,\mathbf{e}^{\text{geo}}(\ell_{i-1},\ell_i)\big] + \mathbf{e}^{\text{pos}}_i .
\end{equation}
We concatenate rather than sum because the three groups live on different scales and have different widths; the projection $\mathbf{W}_{\text{in}}$ can then learn how much of each to keep, whereas addition would force them onto one shared offset space before the model has had any say in the matter.

The components have distinct roles. $\mathbf{e}_{p_i}$ is a learnable POI embedding that absorbs co-visit and category patterns, with index $0$ reserved for padding and held at zero. $\mathbf{e}^{\text{pos}}_i$ is a learned positional embedding over the truncation window $L$, so extrapolation beyond $L$ is never required. $\mathbf{e}^{\text{time}}(t_i)$ concatenates four periodic features, the sine and cosine of local hour-of-day and of day-of-week, with a learnable embedding of a four-way coarse time slot (morning, afternoon, evening, night); the periodic pair captures the smooth daily and weekly rhythm while the slot embedding gives the model a discrete handle on the parts of the day that behave differently. $\mathbf{e}^{\text{geo}}(\ell_{i-1},\ell_i)$ is a two-layer MLP over the \emph{displacement} from the previous check-in, namely $\log(1+\delta)$ for the Haversine distance $\delta$, a learnable embedding of an eight-way distance bucket, and the raw latitude and longitude differences; the first position takes zero displacement. We encode displacement rather than absolute coordinates because absolute position is already carried by the POI embedding, whereas the size and direction of the last move is what distinguishes a stay from a commute. Candidates use the same POI table as the history, so a candidate and its past occurrences share one representation, which keeps the reader and the revisit gate consistent (Section~\ref{sec:scoring}).

A padding mask $\mathbf{M}\in\{0,1\}^{L}$ with 1 at padded positions is applied via each downstream attention call. Left-padding keeps the most recent visit at position $L$ for each user, which makes recency-based indexing trivial and avoids a per-user offset. Before candidate conditioning, CaST-POI applies a stack of causal self-attention blocks over the history embeddings,
\begin{equation}
\mathbf{X} \leftarrow \textsc{HistorySelfAttn}(\mathbf{X}; \mathbf{M}),
\qquad \mathbf{X}\in\mathbb{R}^{L\times d},
\end{equation}
which lets history tokens exchange information with each other before they are read by candidate queries. Masked logits use a large negative constant rather than $-\infty$: where causal masking meets left padding, a query row inside the padded region would otherwise have every key masked and produce a NaN gradient. Outputs at padded positions are zeroed so that padded tokens cannot leak through the residual connections.

\subsection{Candidate-Conditioned Reader}
\label{sec:42}
The core of CaST-POI is multi-head candidate-conditioned cross-attention, in which candidates act as queries and historical visits act as keys and values. This is the standard target-attention pattern from CTR, instantiated with the Transformer scaled dot-product form. For head $h\in\{1,\ldots,H\}$, candidate states are projected to queries $\mathbf{Q}^{(h)}\in\mathbb{R}^{C\times d_h}$ and history states to keys and values $\mathbf{K}^{(h)},\mathbf{V}^{(h)}\in\mathbb{R}^{L\times d_h}$ with $d_h=d/H$, and
\begin{equation}
\label{eq:ccattn}
\textsc{CCAttn}^{(h)} = \text{softmax}\!\left(\frac{\mathbf{Q}^{(h)}\mathbf{K}^{(h)\top}}{\sqrt{d_h}} + \mathbf{B} + \mathbf{M}'\right)\mathbf{V}^{(h)},
\end{equation}
where $\mathbf{B}\in\mathbb{R}^{C\times L}$ is the candidate-relative bias defined in Section~\ref{sec:43} and $\mathbf{M}'$ expresses the padding mask additively, placing the same large negative constant at padded key positions and $0$ elsewhere. The head outputs are concatenated and projected back to width $d$, and we write $\textsc{CCAttn}(\mathbf{U},\mathbf{X},\mathbf{X};\mathbf{B},\mathbf{M})$ for this multi-head result. Multiple heads let different subspaces pick up different facets of the candidate-history interaction, for example spatial proximity in some heads and category affinity in others; the bias $\mathbf{B}$ is shared across heads, since it carries a single signal.
We stack $N$ candidate-conditioned encoder blocks. Each block applies cross-attention and a feed-forward sublayer in sequence, each wrapped by a residual connection and a LayerNorm,
\begin{align}
\mathbf{U}^{(0)} &= \mathbf{E}_C, \\
\widetilde{\mathbf{U}}^{(n)} &= \textsc{LN}_1\big(\mathbf{U}^{(n)} + \textsc{CCAttn}(\mathbf{U}^{(n)},\mathbf{X},\mathbf{X};\mathbf{B},\mathbf{M})\big), \\
\mathbf{U}^{(n+1)} &= \textsc{LN}_2\big(\widetilde{\mathbf{U}}^{(n)} + \textsc{FFN}(\widetilde{\mathbf{U}}^{(n)})\big),
\end{align}
with $\mathbf{E}_C\in\mathbb{R}^{C\times d}$ the candidate embeddings entering at layer $0$ and FFN a two-layer network with ReLU activation and dropout. The residuals preserve the candidate embedding across layers, so a deeper stack does not wash out the identity of the candidate being scored. The final candidate-specific user representation is $\mathbf{z}_u^{(c_j)}=\mathbf{U}^{(N)}_{j,:}$.

A small head then turns this representation into a scalar. We use a two-layer MLP with GELU activation and dropout over the concatenation of three signals, the candidate-specific user representation, the candidate embedding itself, and their element-wise product,
\begin{equation}
\label{eq:refine}
r(u,c_j)=\text{MLP}\big([\mathbf{z}_u^{(c_j)};\,\mathbf{e}_{c_j};\,\mathbf{z}_u^{(c_j)}\odot \mathbf{e}_{c_j}]\big).
\end{equation}
The element-wise product captures multiplicative interactions that concatenation alone cannot express, and the three-signal combination is a standard recipe in the target-attention literature. Note that $r(u,c_j)$ is a \emph{refinement} term rather than the final score: Section~\ref{sec:scoring} adds it to a candidate-agnostic base and a revisit term.

The dominant inference cost is the cross-attention itself, vectorised across candidates, with complexity $\mathcal{O}(N H C L d_h)=\mathcal{O}(N C L d)$ per instance. History keys and values are computed once and reused across every candidate, so the trajectory is never re-encoded per candidate. At full-vocabulary scale the bias tensor would not fit in memory when a whole batch is scored against $C=|\mathcal{P}|$, so candidates are scored in chunks of $1{,}024$ against the cached keys and values, which is a memory-blocking detail that leaves the computed scores identical. What matters is that the cost is one encode plus a batched attention per chunk, rather than the $C$-fold loop a naive candidate-conditioned reader would require, and this is what keeps the ranker single-stage over a vocabulary of ten thousand POIs. Algorithm~\ref{alg:cast_poi_short} implements the forward pass.

\renewcommand{\algorithmicrequire}{\textbf{Input:}}
\renewcommand{\algorithmicensure}{\textbf{Output:}}

\begin{algorithm}[t]
\caption{CaST-POI: Candidate-Conditioned Spatiotemporal Ranking}
\label{alg:cast_poi_short}
\small
\begin{algorithmic}[1]
\REQUIRE history $\mathcal{H}$ with locations $\ell_i$; candidates $\mathcal{C}=\{(c_j,\ell_{c_j})\}_{j=1}^{C}$
\ENSURE score vector $\mathbf{s}\in\mathbb{R}^{C}$

\STATE $\mathcal{H}^{L} \leftarrow$ last $L$ visits;\ \ $\mathcal{H}^{R} \leftarrow$ last $512$ visits \COMMENT{two windows}
\STATE $\mathbf{x}_i \leftarrow \mathbf{W}_{\text{in}}[\mathbf{e}_{p_i};\mathbf{e}^{\text{time}}(t_i);\mathbf{e}^{\text{geo}}(\ell_{i-1},\ell_i)] + \mathbf{e}^{\text{pos}}_i,\ \ (p_i,t_i)\in\mathcal{H}^{L}$
\STATE $\mathbf{X} \leftarrow \textsc{HistorySelfAttn}(\mathbf{x}_{1:L};\mathbf{M})$;\ \ $\mathbf{h}_u \leftarrow \mathbf{X}_{L}$
\STATE $\rho_u \leftarrow \textsc{RevisitGate}(\mathbf{h}_u,\mathcal{H}^{R})$

\STATE $\Delta t_i \leftarrow \max(0,\,t_L - t_i),\ \ i=1,\ldots,L$
\STATE $b_t(\Delta t_i) \leftarrow \mathbf{w}_t^\top\mathbf{g}_t(\textsc{Bkt}_t(\Delta t_i)),\ \ i=1,\ldots,L$
\STATE $b_s(\ell_i,\ell_{c_j}) \leftarrow \mathbf{w}_s^\top\mathbf{g}_s(\textsc{Bkt}_s(\textsc{Hav}(\ell_i,\ell_{c_j}))),\ \ \forall i,j$
\STATE $\mathbf{B}_{j,i} \leftarrow b_t(\Delta t_i) + b_s(\ell_i,\ell_{c_j}),\ \ \forall i,j$ \COMMENT{candidate-relative}

\STATE $\mathbf{U}^{(0)} \leftarrow [\mathbf{e}_{c_1},\ldots,\mathbf{e}_{c_C}]$
\FOR{$n=0$ \TO $N-1$}
    \STATE $\widetilde{\mathbf{U}}^{(n)} \leftarrow \textsc{LN}_1(\mathbf{U}^{(n)} + \textsc{CCAttn}(\mathbf{U}^{(n)},\mathbf{X},\mathbf{X};\mathbf{B},\mathbf{M}))$
    \STATE $\mathbf{U}^{(n+1)} \leftarrow \textsc{LN}_2(\widetilde{\mathbf{U}}^{(n)} + \textsc{FFN}(\widetilde{\mathbf{U}}^{(n)}))$
\ENDFOR

\STATE $r_j \leftarrow \textsc{MLP}([\mathbf{U}^{(N)}_{j};\,\mathbf{e}_{c_j};\,\mathbf{U}^{(N)}_{j}\odot\mathbf{e}_{c_j}]),\ \ j=1,\ldots,C$
\STATE $s_j \leftarrow \mathbf{h}_u^{\top}\mathbf{e}_{c_j} + \beta_{c_j} + \gamma\, r_j + \rho_u(c_j)$ \COMMENT{Eq.~\eqref{eq:score}}
\RETURN $\mathbf{s}=(s_1,\ldots,s_C)$
\end{algorithmic}
\end{algorithm}

\subsection{Spatiotemporal Biases}
\label{sec:43}
The two biases added to the attention logits are where POI-specific geometry enters the model. Both follow the same bucketise-then-embed-then-project design and produce a scalar per query-key pair that is added to the attention logits before the softmax of Eq.~\eqref{eq:ccattn}. Bucketisation lets each bias represent non-monotone preference shifts, for example a sharp fall-off beyond walking distance but a flatter response within it, which a continuous parametric form would have to approximate. Learnable per-bucket embeddings allow each bucket to acquire its own offset without imposing a fixed functional form on the decay, and the linear projection to a scalar keeps the parameter overhead negligible relative to the attention weights.

Recency is measured relative to the last timestamp in the sequence,
\begin{equation}
\Delta t_i = \max(0,\, t_{L} - t_i),
\end{equation}
so that $\Delta t_i\ge0$ by construction and any padded position collapses into a gap that the padding mask then zeroes out in the attention weights. Measuring recency against the last visit rather than an absolute clock makes the bias user-agnostic and stable under trajectory truncation.

\paragraph{\textbf{Temporal bias}}
$\Delta t_i$ is bucketised into six logarithmic ranges delimited by the five boundaries $\{1\text{h},\,6\text{h},\,24\text{h},\,7\text{d},\,30\text{d}\}$, so the lowest range is $[0,1\text{h})$ and the highest is $[30\text{d},\infty)$. The boundaries align with human mobility periods, which keeps the temporal resolution fine where it matters for next-visit decisions and coarse where it does not. Each bucket has a learnable embedding followed by a linear projection to a scalar,
\begin{equation}
b_t(\Delta t_i) = \mathbf{w}_t^\top \mathbf{g}_t\big(\text{bucket}(\Delta t_i)\big),
\end{equation}
giving one scalar per history position, broadcast across all $C$ candidates, because the recency of a past visit does not depend on which candidate is being scored.

\paragraph{\textbf{Spatial bias}}
The spatial bias is where candidate dependence enters the model explicitly. For every pair of history location $\ell_i$ and candidate location $\ell_{c_j}$ we compute the Haversine distance in kilometres, bucketise it into six ranges with boundaries $\{0.5,\,2,\,5,\,20,\,100\}$ km, and apply the same embed-then-project scheme,
\begin{equation}
b_s(\ell_i,\ell_{c_j}) = \mathbf{w}_s^\top \mathbf{g}_s\big(\text{bucket}(\textsc{Hav}(\ell_i,\ell_{c_j}))\big).
\end{equation}
We use Haversine rather than Euclidean distance on raw coordinates because the datasets span regions where the flat-Earth approximation breaks at long range. The boundaries track walking distance, same-neighbourhood, same-district, within-city and inter-city scales, which correspond to the spatial categories mobility decisions actually distinguish between; the top bucket matters for CA in particular, which spans a whole state rather than one city. The result is $\mathbf{B}_s\in\mathbb{R}^{C\times L}$, a separate bias vector for every candidate.

\paragraph{\textbf{Combined bias}}
The bias added to the attention logits is the position-wise sum of the two terms,
\begin{equation}
\mathbf{B}_{j,i} = b_t(\Delta t_i) + b_s(\ell_i,\ell_{c_j}),
\end{equation}
so that $\mathbf{B}\in\mathbb{R}^{C\times L}$ matches the shape of the attention logits. Additive combination lets the two biases compose independently: a history position that is both recent and spatially close to the candidate receives the sum of both preferred offsets, while a position that is recent but far away sees recency alone in the weighting.

\subsection{Scoring: Base, Refinement, and Revisit Gate}
\label{sec:scoring}
The candidate-conditioned reader of Section~\ref{sec:42} supplies one of four additive terms. Writing $\mathbf{h}_u$ for the backbone state at the last history position, the score of candidate $c$ for user $u$ is
\begin{equation}
\label{eq:score}
s(u,c) = \underbrace{\mathbf{h}_u^{\top}\mathbf{e}_{c} + \beta_c}_{\text{candidate-agnostic base}}
\;+\; \underbrace{\gamma\, r(u,c)}_{\text{candidate-conditioned}}
\;+\; \underbrace{\rho_u(c)}_{\text{revisit gate}},
\end{equation}
where $\beta_c$ is a per-POI scalar bias, $r(u,c)$ is the refinement of Eq.~\eqref{eq:refine}, and $\gamma$ is a single learnable scalar initialised to $0$.

Each term is there for a reason. The dot-product base $\mathbf{h}_u^{\top}\mathbf{e}_c$ is exactly the scoring rule of a SASRec-style sequential recommender, so CaST-POI starts from a configuration at least as expressive as that baseline and the candidate-conditioned branch is measured as a \emph{marginal} gain over it rather than against a weak scorer. Initialising $\gamma$ at $0$ makes this explicit: at the first optimisation step the candidate-conditioned branch contributes nothing to the score, and any weight it acquires is weight the optimiser chose to give it. The per-POI bias $\beta_c$ absorbs global popularity so that the attention machinery does not have to spend capacity on it.

\paragraph{\textbf{Revisit gate}}
The last term is an explicit channel for revisit evidence, motivated by how far check-in benchmarks are dominated by repetition: $64.1\%$ of NYC test targets, $68.5\%$ of TKY targets and $44.7\%$ of CA targets are places the user has already visited. From the visible history we derive three per-candidate features, $\log(1+\text{cnt}_u(c))$ for the number of past visits $\text{cnt}_u(c)$ to $c$, a recency score $\text{rec}_u(c)$ equal to the normalised position of the most recent visit to $c$, and a binary visited indicator. These features are counted directly from the history rather than learned; the only parameters involved are those of a small MLP that maps the backbone state $\mathbf{h}_u$ to three mixing weights $\mathbf{w}(u)\in\mathbb{R}^{3}$, and the gate is their inner product with the feature triple,
\begin{equation}
\rho_u(c) = \mathbf{w}(u)^{\top}\big[\log(1+\text{cnt}_u(c)),\; \text{rec}_u(c),\; \mathbf{1}[\text{cnt}_u(c)>0]\big].
\end{equation}
Conditioning the weights on the user state rather than fixing them lets the gate be turned down for a user whose trajectory looks exploratory. The revisit features are counted over a longer window than the attention window: the reader truncates to the last $L=50$ visits because attention costs $\mathcal{O}(L)$ per candidate, but counting does not, so the gate sees up to $512$ past check-ins. Restricting it to the $50$-visit reader window would make a target look like a first visit when it is not, with the fraction of targets visible as revisits falling to $62.0\%$, $63.2\%$ and $42.9\%$ respectively.

\subsection{Optimization}
\label{sec:optim}
Training scores the positive against the \emph{entire} POI vocabulary, which is the same $|\mathcal{P}|$-way problem the model is asked to solve at evaluation time. This matters more than it may appear. A sampled softmax with a few hundred negatives trains a discriminator over a far smaller space than the $5{,}000$- to $10{,}000$-way problem posed at test time, and the sequential baselines we compare against are all trained with full-vocabulary cross-entropy, so a sampled objective on our side alone would have handicapped our own model relative to theirs. The vocabulary is small enough that a dense softmax is affordable, and the revisit gate already materialises a dense $|\mathcal{P}|$-wide tensor, so the full-width computation is paid for either way.

The loss combines a smoothed cross-entropy with a hard-negative ranking margin,
\begin{equation}
\mathcal{L} = \textsc{CE}_{\epsilon}(\mathbf{s}, y)
\;+\; \lambda \cdot \frac{1}{k}\sum_{c \in \mathcal{N}_k(u)} \big[\, m - (s_y - s_c) \,\big]_{+},
\end{equation}
where $\mathbf{s}\in\mathbb{R}^{|\mathcal{P}|}$ collects the per-candidate logits, $y$ is the target POI, $\epsilon$ is the label-smoothing factor, $m$ is the margin, and $\mathcal{N}_k(u)$ is the set of the $k$ highest-scoring non-target POIs for that instance. The cross-entropy term calibrates the whole distribution, while the margin term acts only on the handful of items actually competing with the target, which is where the ranking metrics are decided. Label smoothing matters here because with full-vocabulary scoring the negative set necessarily contains other users' targets, so a hard zero on every non-target is the wrong supervision signal.

Check-in datasets are heavily biased toward revisits, so a ranker trained on them can collapse into predicting history membership and ignore genuinely new locations. We therefore upweight \textit{explore} instances, in which the target POI appears nowhere in the user's history, by a constant factor $w_{\text{explore}}>1$ on the per-instance loss. This pushes the spatial bias to be useful for unfamiliar regions rather than only for revisits, and it is the counterweight to the revisit gate of Section~\ref{sec:scoring}: the gate makes revisit prediction easy, and the reweighting stops that from being the only thing the model learns.

\section{Performance Evaluation}

\subsection{Experimental Setup}

\subsubsection{Datasets}

We evaluate CaST-POI on three public benchmarks: NYC and TKY from Foursquare~\cite{yang2015modeling}, and CA from Gowalla~\cite{yuan2013time}. Rather than run our own filtering, we start from the published preprocessed files of the STHGCN/LLM4POI line of work~\cite{li2024large}, whose $k$-core step and trajectory segmentation are already baked in, and we verified that our copy is byte-identical to the released files. Statistics are given in Table~\ref{tab:dataset}. The datasets cover different mobility regimes: NYC and TKY are dense metropolitan areas where many POIs lie within walking distance and candidates in a small radius often share categories, whereas CA spans a whole state with sparser POIs and automobile-scale travel, where coarse spatial preferences dominate. On top of these files we build a standard per-user leave-one-out split~\cite{kang2018self}, taking each user's last check-in as the test target and the second-to-last for validation, so that every user contributes exactly one test instance. The split is content-hashed and its fingerprint is checked at load time by every model we run, so all rows of Table~\ref{tab:main} are computed on a bit-identical split; the split script and its integrity checks are released with our code. Every learned model is trained with three random seeds.

\begin{table}[h]
\centering
\caption{Dataset statistics under the per-user leave-one-out split. \#POIs counts POI ids that occur in the split; \#Test is the number of users retained after the unseen-POI filter. Avg.\ Len.\ is check-ins per user.}
\label{tab:dataset}
\resizebox{\linewidth}{!}{
\begin{tabular}{lccccc}
\toprule
\textbf{Dataset} & \textbf{\#Users} & \textbf{\#POIs} & \textbf{\#Check-ins} & \textbf{Avg. Len.} & \textbf{\#Test} \\
\midrule
NYC & 1,047 & 4,974  & 101,753 & 97.2  & 1,044 \\
TKY & 2,280 & 7,832  & 403,146 & 176.8 & 2,280 \\
CA  & 3,956 & 9,689  & 221,717 & 56.0  & 3,956 \\
\bottomrule
\end{tabular}
}
\end{table}

\begin{table}[t]
\centering
\caption{Overall comparison under per-user leave-one-out with full-vocabulary ranking (mean $\pm$ std over three seeds). The best value per column is in \textbf{bold}, and the strongest baseline, against which \textit{Improv.} is computed, is \underline{underlined}. Significance is a paired per-sample Wilcoxon test Holm-corrected across the 15 headline comparisons ($^{*}p<0.05$, $^{**}p<0.01$, $^{***}p<0.001$). The revisit heuristic is deterministic and has no std.}
\label{tab:main}
\scriptsize
\setlength{\tabcolsep}{2.4pt}
\newcommand{\std}[1]{{\tiny$\pm$#1}}
\resizebox{\columnwidth}{!}{
\begin{tabular}{l|ccccc}
\toprule
\multicolumn{6}{c}{\textbf{NYC (\%)}} \\
\midrule
\textbf{Method} & \textbf{HR@5$\uparrow$} & \textbf{HR@10$\uparrow$} & \textbf{NDCG@5$\uparrow$} & \textbf{NDCG@10$\uparrow$} & \textbf{MRR$\uparrow$} \\
\midrule
SASRec~\cite{kang2018self}         & 50.42\std{0.06} & 56.99\std{0.10} & 38.01\std{0.12} & 40.14\std{0.13} & 35.35\std{0.17} \\
BERT4Rec~\cite{sun2019bert4rec}    & 47.67\std{0.45} & 54.69\std{0.25} & 36.14\std{0.06} & 38.43\std{0.10} & 33.78\std{0.13} \\
Caser~\cite{tang2018personalized}  & 48.12\std{0.31} & 54.15\std{0.06} & 35.84\std{0.18} & 37.80\std{0.15} & 33.12\std{0.17} \\
NARM~\cite{li2017neural}           & 49.39\std{0.53} & 55.87\std{0.22} & 37.15\std{0.38} & 39.30\std{0.27} & 34.55\std{0.27} \\
SR-GNN~\cite{wu2019session}        & 44.06\std{0.48} & 52.81\std{0.11} & 32.82\std{0.27} & 35.67\std{0.18} & 30.90\std{0.27} \\
STAMP~\cite{liu2018stamp}          & 41.60\std{1.21} & 46.33\std{2.13} & 32.44\std{0.76} & 33.98\std{1.02} & 30.51\std{0.64} \\
CORE~\cite{hou2022core}            & \textbf{51.31\std{0.11}} & \underline{57.76\std{0.17}} & \underline{38.30\std{0.18}} & \underline{40.40\std{0.09}} & \underline{35.38\std{0.16}} \\
Revisit heuristic                  & 41.19\phantom{\std{0.00}} & 54.60\phantom{\std{0.00}} & 28.66\phantom{\std{0.00}} & 33.05\phantom{\std{0.00}} & 26.94\phantom{\std{0.00}} \\
\midrule
\rowcolor{lightblue}\textbf{CaST-POI} & 51.02\std{0.24} & \textbf{58.01\std{0.15}} & \textbf{39.51\std{0.22}} & \textbf{41.80\std{0.17}} & \textbf{37.20\std{0.16}} \\
\rowcolor{lightblue}\textit{Improv.} & $-0.57\%$ & $+0.43\%$ & $+3.16\%$ & $+3.47\%$ & $+5.14\%$ \\
\midrule
\multicolumn{6}{c}{\textbf{TKY (\%)}} \\
\midrule
\textbf{Method} & \textbf{HR@5$\uparrow$} & \textbf{HR@10$\uparrow$} & \textbf{NDCG@5$\uparrow$} & \textbf{NDCG@10$\uparrow$} & \textbf{MRR$\uparrow$} \\
\midrule
SASRec~\cite{kang2018self}         & \underline{46.61\std{0.77}} & \underline{56.02\std{0.41}} & 34.09\std{0.69} & \underline{37.14\std{0.60}} & 32.12\std{0.67} \\
BERT4Rec~\cite{sun2019bert4rec}    & 44.28\std{0.94} & 53.67\std{0.98} & 33.40\std{0.87} & 36.46\std{0.80} & 31.96\std{0.84} \\
Caser~\cite{tang2018personalized}  & 43.60\std{0.71} & 51.81\std{0.46} & 33.18\std{0.96} & 35.86\std{0.87} & 31.69\std{1.00} \\
NARM~\cite{li2017neural}           & 45.04\std{0.24} & 53.55\std{0.08} & 34.15\std{0.33} & 36.92\std{0.27} & \underline{32.57\std{0.34}} \\
SR-GNN~\cite{wu2019session}        & 45.44\std{0.29} & 53.49\std{0.53} & \underline{34.15\std{0.35}} & 36.74\std{0.45} & 32.33\std{0.41} \\
STAMP~\cite{liu2018stamp}          & 42.87\std{0.44} & 49.82\std{0.29} & 32.91\std{0.45} & 35.17\std{0.31} & 31.40\std{0.40} \\
CORE~\cite{hou2022core}            & 43.41\std{0.71} & 53.48\std{0.89} & 29.77\std{0.40} & 33.05\std{0.42} & 27.58\std{0.27} \\
Revisit heuristic                  & 29.25\phantom{\std{0.00}} & 41.54\phantom{\std{0.00}} & 19.29\phantom{\std{0.00}} & 23.22\phantom{\std{0.00}} & 18.97\phantom{\std{0.00}} \\
\midrule
\rowcolor{lightblue}\textbf{CaST-POI} & \textbf{48.39\std{0.81}} & \textbf{57.76\std{0.75}} & \textbf{36.73\std{0.34}} & \textbf{39.76\std{0.36}} & \textbf{35.05\std{0.22}} \\
\rowcolor{lightblue}\textit{Improv.} & $+3.82\%$$^{*}$ & $+3.11\%$$^{*}$ & $+7.55\%$$^{***}$ & $+7.05\%$$^{***}$ & $+7.61\%$$^{***}$ \\
\midrule
\multicolumn{6}{c}{\textbf{CA (\%)}} \\
\midrule
\textbf{Method} & \textbf{HR@5$\uparrow$} & \textbf{HR@10$\uparrow$} & \textbf{NDCG@5$\uparrow$} & \textbf{NDCG@10$\uparrow$} & \textbf{MRR$\uparrow$} \\
\midrule
SASRec~\cite{kang2018self}         & \underline{35.48\std{0.19}} & \underline{44.25\std{0.23}} & \underline{24.81\std{0.13}} & \underline{27.65\std{0.04}} & \underline{23.69\std{0.08}} \\
BERT4Rec~\cite{sun2019bert4rec}    & 28.99\std{0.35} & 36.80\std{0.48} & 21.64\std{0.32} & 24.18\std{0.39} & 21.37\std{0.34} \\
Caser~\cite{tang2018personalized}  & 27.32\std{0.40} & 34.34\std{0.48} & 20.23\std{0.27} & 22.51\std{0.31} & 19.88\std{0.25} \\
NARM~\cite{li2017neural}           & 32.01\std{0.45} & 40.09\std{0.48} & 23.66\std{0.35} & 26.28\std{0.35} & 23.11\std{0.29} \\
SR-GNN~\cite{wu2019session}        & 30.74\std{0.33} & 38.29\std{0.55} & 22.75\std{0.27} & 25.19\std{0.19} & 22.22\std{0.25} \\
STAMP~\cite{liu2018stamp}          & 27.92\std{1.01} & 34.50\std{1.13} & 21.11\std{0.57} & 23.24\std{0.61} & 20.80\std{0.45} \\
CORE~\cite{hou2022core}            & 31.66\std{0.14} & 39.92\std{0.25} & 23.10\std{0.24} & 25.77\std{0.27} & 22.57\std{0.29} \\
Revisit heuristic                  & 28.64\phantom{\std{0.00}} & 35.52\phantom{\std{0.00}} & 21.13\phantom{\std{0.00}} & 23.35\phantom{\std{0.00}} & 20.05\phantom{\std{0.00}} \\
\midrule
\rowcolor{lightblue}\textbf{CaST-POI} & \textbf{38.17\std{0.75}} & \textbf{46.29\std{0.29}} & \textbf{28.23\std{0.47}} & \textbf{30.86\std{0.30}} & \textbf{27.12\std{0.36}} \\
\rowcolor{lightblue}\textit{Improv.} & $+7.58\%$$^{***}$ & $+4.61\%$$^{**}$ & $+13.78\%$$^{***}$ & $+11.61\%$$^{***}$ & $+14.48\%$$^{***}$ \\
\bottomrule
\end{tabular}
}
\end{table}

\subsubsection{Baselines}

We compare against seven sequential recommenders spanning the main architectural families: the self-attentive rankers SASRec~\cite{kang2018self} and BERT4Rec~\cite{sun2019bert4rec}, the attention-pooling encoders NARM~\cite{li2017neural} and STAMP~\cite{liu2018stamp}, the convolutional Caser~\cite{tang2018personalized}, the graph-based SR-GNN~\cite{wu2019session}, and CORE~\cite{hou2022core}, which constrains session and item representations to a single space. All are trained by us with the official RecBole~\cite{zhao2021recbole} implementations on the identical leave-one-out split and ranked full-vocabulary, so Table~\ref{tab:main} is a paired comparison rather than numbers copied across papers. None of them has a channel for candidate-relative spatial relevance, the property under test. We further include a zero-parameter \textbf{revisit heuristic} that scores each candidate by its visit count in the user's history times its global popularity, since next-POI benchmarks are revisit-dominated and a method that cannot beat this reference has not shown that it models mobility rather than repetition. We do not report published numbers for POI-specific and LLM-based systems such as GETNext, MTNet, LLM4POI or CoMaPOI, because they are obtained under a sampled-candidate protocol on a different split.

\subsubsection{Evaluation Protocol}

We report Hit Rate (\textbf{HR@$k$}), NDCG@$k$ and Mean Reciprocal Rank (\textbf{MRR}) with $k\in\{5,10\}$, ranking every test instance against the \textbf{full POI vocabulary} of its dataset ($4{,}974$ / $7{,}832$ / $9{,}689$ items) rather than a sampled pool. Section~\ref{sec:candidate_pool_size} shows why: at a pool of 100 the ordering of methods on CA reverses relative to full ranking, so a sampled protocol can rank a worse model first. Ties are broken by the mid-rank convention, $\text{rank}=1+\#\{\text{strictly better}\}+\#\{\text{tied}\}/2$, applied identically to every method, which matters because the revisit heuristic scores every unvisited POI exactly zero and would otherwise be handed a top-10 rank on targets it never saw.

\textbf{Significance.} Since all methods are evaluated on the same instances, we test paired per-sample metric values rather than seed means. For each comparison we average the per-sample metric over the three seeds on each side and apply a two-sided Wilcoxon signed-rank test with Holm-Bonferroni correction over the family of tests being made, 15 for Table~\ref{tab:main} and 5 within each dataset-metric cell for Table~\ref{tab:ablation}. A paired bootstrap over instances ($B=20{,}000$), corrected the same way, agrees on 13 of the 15 headline comparisons; it rejects on NYC NDCG@10 and MRR where the Wilcoxon test does not, since the bootstrap tests the mean difference while the signed-rank test tests a symmetry hypothesis about the per-instance differences. We mark the tables with the Wilcoxon result because it is the conservative one, and discuss the disagreement in Section~\ref{sec:comparison}.

\subsubsection{Implementation Details}

CaST-POI uses embedding dimension $d=128$, 4 attention heads, 2 causal self-attention layers and 2 candidate-conditioned cross-attention layers, with histories truncated to $L=50$ visits for the reader and 512 for the revisit gate, and 6 buckets each for the temporal and spatial biases as specified in Section~\ref{sec:43}. Training uses AdamW (learning rate $10^{-3}$, weight decay $10^{-4}$), batch size 128, dropout $0.1$, gradient clipping at $5.0$, label smoothing $\epsilon=0.02$, margin weight $\lambda=0.5$ with $m=1.0$ over the top $k=10$ hard negatives, and explore weight $w_{\text{explore}}=1.5$, for up to 50 epochs with 3 warm-up epochs, cosine decay and early stopping on validation HR@10 with patience 10. One configuration is used for all three datasets and nothing is tuned per dataset. The model has $1.13$M, $1.49$M and $1.73$M parameters on NYC, TKY and CA, the difference being the POI embedding table, and all experiments run on a single NVIDIA A100.

\subsection{Performance Comparison}
\label{sec:comparison}

\textbf{Comparison across Datasets.}
Table~\ref{tab:main} reports the overall comparison. CaST-POI holds the best value in 14 of the 15 cells, the exception being HR@5 on NYC where CORE is ahead by $0.29$ points. The gains are largest on CA, the geographically dispersed dataset ($+14.48\%$ MRR and $+13.78\%$ NDCG@5 over SASRec), and smallest on NYC. This ordering is the opposite of what a dense-urban story would predict, and it is the expected one for our mechanism: a state-scale region contains candidates hundreds of kilometres from anything the user has visited, whereas in Manhattan almost every candidate is within a few kilometres of almost every past visit, leaving the spatial bias little to separate.

\textbf{Where is the gain from?} 
The improvement is consistently larger on the position-sensitive metrics than on HR@$k$. On NYC, HR@10 improves by $0.43\%$ against $5.14\%$ for MRR; on TKY the corresponding figures are $3.11\%$ and $7.61\%$. CaST-POI is therefore mostly \emph{reordering} items that a strong sequential ranker already retrieves rather than retrieving items those rankers miss, which is what candidate conditioning is designed to do. On NYC none of the five improvements survives Holm correction under the Wilcoxon test reported in Table~\ref{tab:main}, and on HR@$k$ the point estimates sit inside the seed-to-seed noise, so CaST-POI and CORE are indistinguishable there under either test. The two position-sensitive metrics are borderline: a paired bootstrap on the mean difference rejects on NDCG@10 and MRR after the same correction ($95\%$ CIs $[+0.31,+2.49]$ and $[+0.59,+3.06]$) while the signed-rank test does not. We take the conservative reading and call NYC a null result, noting that it is also the smallest test set ($n=1{,}044$) and so has the least power of the three.

\textbf{Comparison across baselines.}
No single baseline is strongest everywhere. CORE leads on NYC but has the worst MRR of any neural baseline on TKY ($27.58$ against SASRec's $32.12$), as its consistent-representation constraint helps when sessions are short and hurts when they are long ($176.8$ average check-ins per user on TKY against $97.2$ on NYC), which is why we compare against the per-dataset best rather than a fixed reference. The zero-parameter revisit heuristic is a sharper diagnostic: on NYC it reaches HR@10 $54.60$, beating SR-GNN, STAMP and Caser, so on this benchmark a published neural ranker can be outperformed by counting. It falls well behind the leaders on TKY ($41.54$ against $56.02$) and CA ($35.52$ against $44.25$), where longer trajectories or a larger region make pure repetition insufficient.

\begin{table}[t]
\centering
\caption{Component ablation (mean $\pm$ sample std over three seeds). Superscripts mark the significance of the drop from the full model under a paired Wilcoxon test, Holm-corrected across the five variants within each dataset and metric ($^{*}p<0.05$, $^{**}p<0.01$, $^{***}p<0.001$).}
\label{tab:ablation}
\scriptsize
\setlength{\tabcolsep}{2pt}
\newcommand{\std}[1]{{\tiny$\pm$#1}}
\resizebox{\columnwidth}{!}{
\begin{tabular}{l|ccccc}
\toprule
\multicolumn{6}{c}{\textbf{NYC (\%)}} \\
\midrule
\textbf{Configuration} & \textbf{HR@5$\uparrow$} & \textbf{HR@10$\uparrow$} & \textbf{NDCG@5$\uparrow$} & \textbf{NDCG@10$\uparrow$} & \textbf{MRR$\uparrow$} \\
\midrule
\rowcolor{lightblue}\textbf{Full Model} & 50.48\std{0.35}\phantom{$^{***}$} & 57.85\std{0.86}\phantom{$^{***}$} & 39.17\std{0.37}\phantom{$^{***}$} & 41.58\std{0.07}\phantom{$^{***}$} & 36.97\std{0.23}\phantom{$^{***}$} \\
w/o Temporal Bias & 50.29\std{0.00}\phantom{$^{***}$} & 57.79\std{0.31}\phantom{$^{***}$} & 38.94\std{0.32}\phantom{$^{***}$} & 41.37\std{0.27}\phantom{$^{***}$} & 36.78\std{0.36}\phantom{$^{***}$} \\
w/o Spatial Bias & 49.65\std{0.43}\phantom{$^{***}$} & 56.74\std{0.87}$^{*}$ & 38.36\std{0.99}$^{*}$ & 40.66\std{1.04}$^{**}$ & 36.08\std{1.25}$^{**}$ \\
w/o Cand.\ Conditioning & 49.87\std{1.43}\phantom{$^{***}$} & 57.02\std{0.58}\phantom{$^{***}$} & 37.94\std{2.28}$^{**}$ & 40.24\std{1.97}$^{***}$ & 35.44\std{2.36}$^{***}$ \\
w/o Revisit Gate & 48.24\std{0.45}$^{***}$ & 54.53\std{0.15}$^{***}$ & 37.48\std{0.46}$^{**}$ & 39.53\std{0.41}$^{***}$ & 35.40\std{0.50}$^{*}$ \\
w/o Backbone & 50.61\std{0.58}\phantom{$^{***}$} & 57.47\std{0.25}\phantom{$^{***}$} & 38.16\std{0.98}$^{*}$ & 40.41\std{0.86}$^{**}$ & 35.61\std{1.03}$^{*}$ \\
\midrule
\multicolumn{6}{c}{\textbf{TKY (\%)}} \\
\midrule
\textbf{Configuration} & \textbf{HR@5$\uparrow$} & \textbf{HR@10$\uparrow$} & \textbf{NDCG@5$\uparrow$} & \textbf{NDCG@10$\uparrow$} & \textbf{MRR$\uparrow$} \\
\midrule
\rowcolor{lightblue}\textbf{Full Model} & 48.17\std{0.42}\phantom{$^{***}$} & 57.82\std{0.45}\phantom{$^{***}$} & 36.64\std{0.24}\phantom{$^{***}$} & 39.78\std{0.15}\phantom{$^{***}$} & 35.09\std{0.16}\phantom{$^{***}$} \\
w/o Temporal Bias & 48.49\std{1.13}\phantom{$^{***}$} & 57.57\std{0.85}\phantom{$^{***}$} & 36.77\std{1.42}\phantom{$^{***}$} & 39.73\std{1.32}\phantom{$^{***}$} & 35.06\std{1.54}\phantom{$^{***}$} \\
w/o Spatial Bias & 47.44\std{0.38}$^{*}$ & 56.62\std{0.63}$^{**}$ & 36.31\std{0.40}\phantom{$^{***}$} & 39.30\std{0.48}\phantom{$^{***}$} & 34.77\std{0.44}\phantom{$^{***}$} \\
w/o Cand.\ Conditioning & 47.22\std{0.77}$^{*}$ & 55.73\std{1.17}$^{***}$ & 36.13\std{0.34}\phantom{$^{***}$} & 38.88\std{0.30}$^{*}$ & 34.53\std{0.31}$^{*}$ \\
w/o Revisit Gate & 46.67\std{0.56}$^{***}$ & 54.75\std{1.01}$^{***}$ & 36.15\std{0.54}\phantom{$^{***}$} & 38.79\std{0.69}$^{*}$ & 34.72\std{0.62}\phantom{$^{***}$} \\
w/o Backbone & 48.77\std{0.47}\phantom{$^{***}$} & 58.06\std{0.49}\phantom{$^{***}$} & 37.08\std{0.58}\phantom{$^{***}$} & 40.09\std{0.67}\phantom{$^{***}$} & 35.39\std{0.72}\phantom{$^{***}$} \\
\midrule
\multicolumn{6}{c}{\textbf{CA (\%)}} \\
\midrule
\textbf{Configuration} & \textbf{HR@5$\uparrow$} & \textbf{HR@10$\uparrow$} & \textbf{NDCG@5$\uparrow$} & \textbf{NDCG@10$\uparrow$} & \textbf{MRR$\uparrow$} \\
\midrule
\rowcolor{lightblue}\textbf{Full Model} & 37.54\std{0.50}\phantom{$^{***}$} & 45.95\std{0.72}\phantom{$^{***}$} & 28.08\std{0.21}\phantom{$^{***}$} & 30.80\std{0.30}\phantom{$^{***}$} & 27.17\std{0.23}\phantom{$^{***}$} \\
w/o Temporal Bias & 37.55\std{0.40}\phantom{$^{***}$} & 45.97\std{0.80}\phantom{$^{***}$} & 28.01\std{0.40}\phantom{$^{***}$} & 30.74\std{0.54}\phantom{$^{***}$} & 27.09\std{0.51}\phantom{$^{***}$} \\
w/o Spatial Bias & 35.33\std{0.79}$^{***}$ & 43.07\std{1.64}$^{***}$ & 26.39\std{0.57}$^{***}$ & 28.89\std{0.79}$^{***}$ & 25.48\std{0.62}$^{***}$ \\
w/o Cand.\ Conditioning & 35.13\std{0.26}$^{***}$ & 42.91\std{0.12}$^{***}$ & 26.17\std{0.07}$^{***}$ & 28.70\std{0.08}$^{***}$ & 25.25\std{0.12}$^{***}$ \\
w/o Revisit Gate & 35.64\std{0.99}$^{***}$ & 44.29\std{0.78}$^{***}$ & 26.43\std{0.60}$^{***}$ & 29.23\std{0.54}$^{***}$ & 25.75\std{0.44}$^{**}$ \\
w/o Backbone & 36.30\std{0.64}$^{**}$ & 44.92\std{0.44}$^{*}$ & 27.21\std{0.21}$^{***}$ & 30.01\std{0.16}$^{**}$ & 26.48\std{0.08}$^{**}$ \\
\bottomrule
\end{tabular}
}
\end{table}



\begin{figure*}[t]
    \centering
    \begin{subfigure}[t]{0.33\textwidth}
        \centering
        \includegraphics[width=\linewidth]{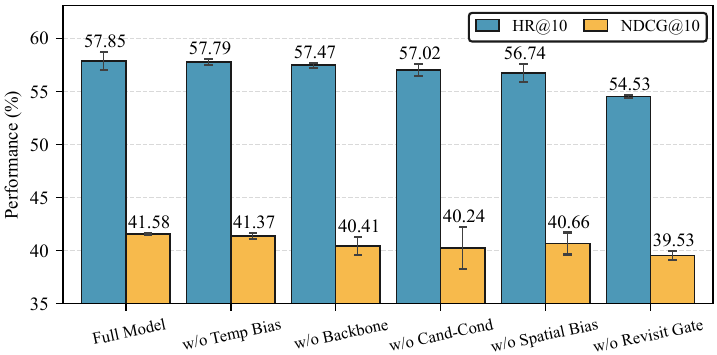}
        \caption{Component ablation on NYC.}
        \label{fig:ablation}
    \end{subfigure}
    \hfill
    \begin{subfigure}[t]{0.24\textwidth}
        \centering
        \includegraphics[width=\linewidth]{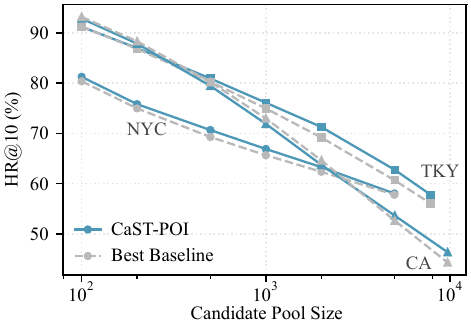}
        \caption{HR@10 vs.\ candidate pool size.}
        \label{fig:candidate_pool}
    \end{subfigure}
    \hfill
    \begin{subfigure}[t]{0.40\textwidth}
        \centering
        \includegraphics[width=\linewidth]{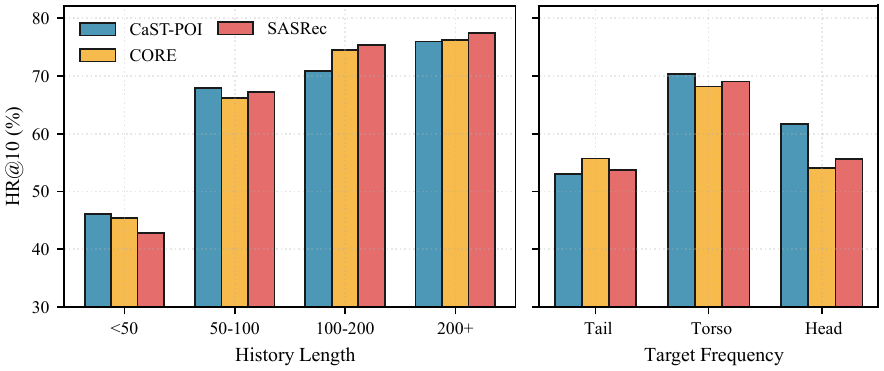}
        \caption{Robustness across strata (NYC).}
        \label{fig:robustness}
    \end{subfigure}
    \caption{Analysis panels. (a) Ablation on NYC, error bars are the sample std over 3 seeds. (b) HR@10 against candidate pool size, computed in closed form from the full-vocabulary ranks (Section~\ref{sec:candidate_pool_size}); solid is CaST-POI, dashed the strongest baseline on that dataset. (c) HR@10 on NYC by user history length and by target-POI training frequency.}
    \label{fig:analysis}
\end{figure*}

\subsection{Ablation Study}
\label{sec:ablation}

Table~\ref{tab:ablation} removes one component at a time and retrains from scratch on all three benchmarks; Figure~\ref{fig:ablation} visualises the NYC column. Each variant is compared against the full model with a paired test, so the question is whether the drop exceeds instance-level noise rather than whether the mean moved. One caveat applies throughout: although we fix the Python, NumPy and Torch seeds and set cuDNN to deterministic mode, the revisit gate builds its per-candidate counts with scatter-add and scatter-reduce, whose GPU atomics are non-deterministic. Two runs of an identical configuration therefore need not agree, and the full model of Table~\ref{tab:ablation} and CaST-POI in Table~\ref{tab:main} are exactly such a pair, differing by up to $0.63$ points. The reported standard deviations mix seed variation with this jitter and should be read as a lower bound on how much a re-run can move, which is also why the tests are paired at the instance level rather than computed over seed means.

\textbf{The spatial bias and the candidate-conditioned reader carry the contribution.}
Removing the candidate-relative spatial bias costs $2.88$ HR@10 on CA, $1.20$ on TKY and $1.11$ on NYC, and removing candidate conditioning entirely costs $3.04$, $2.09$ and $0.83$. Both are significant on every metric on CA and on the hit-rate metrics on TKY. The ordering across datasets matches the main table: in a state-scale region the distance between a candidate and a past visit separates candidates, whereas inside one dense city everything is nearby and it largely does not. A design whose benefit comes from geometry should show exactly this gradient.

\textbf{The temporal bias contributes nothing measurable.}
Removing it changes HR@10 by $-0.06$ on NYC, $-0.25$ on TKY and $+0.02$ on CA. Not one of the fifteen metric-dataset cells is significant, and on CA the ablated model is nominally ahead, so we report this as a negative result rather than absorbing it into a combined ``w/o both biases'' row. The likely reason is structural: as a per-key offset independent of the candidate, the temporal bias can only apply the same reweighting for every candidate inside the softmax over history positions, and recency is already encoded by the causal backbone and the positional embedding. On this evidence the component should be dropped and the claim narrowed to the candidate-relative spatial bias.

\textbf{The revisit gate is the single largest component on the dense datasets.}
Removing it costs $3.32$ HR@10 on NYC and $3.07$ on TKY, more than any other ablation there and significant at $p<0.001$ in both. On CA the ordering flips and the spatial bias and candidate conditioning matter more ($-2.88$ and $-3.04$ against $-1.66$), consistent with CA having the lowest revisit rate of the three ($44.7\%$ of test targets against $64.1\%$ on NYC and $68.5\%$ on TKY). Repetition and geometry act as substitutes: where users mostly return to known places, counting returns is what matters; where they range widely, distance is.

\textbf{The self-attention backbone is not load-bearing on TKY.}
Removing it leaves NYC and CA slightly worse ($-0.38$ and $-1.03$ HR@10) but leaves TKY nominally \emph{better} ($+0.24$, not significant). With the revisit gate and the candidate-conditioned reader both present, the backbone is largely redundant for hit rate on the longest-trajectory dataset, though it still helps the position-sensitive metrics on NYC and CA. We keep it because it makes the candidate-agnostic base of Eq.~\eqref{eq:score} equivalent to a SASRec-style scorer, the reference against which the refinement term is measured.

\subsection{Impact of Candidate Pool Size}
\label{sec:candidate_pool_size}

Much of the POI literature evaluates against a sampled pool of one positive and 99 uniformly drawn negatives, a protocol known to be inconsistent with full ranking~\cite{krichene2020sampled}. Because we retain the full-vocabulary rank $r$ of every test target, we can compute what any method would have scored under that protocol without re-running it: the number of higher-scored items surviving into a pool of $N$ items is hypergeometric, so
\begin{equation}
\Pr[\text{rank}_N \le k] \;=\; F_{\text{hyp}}\big(k-1;\, |\mathcal{P}|-1,\, r-1,\, N-1\big),
\end{equation}
and HR@$k$ at pool size $N$ averages this over test instances. Figure~\ref{fig:candidate_pool} plots the result from $N=100$ to full ranking.

\textbf{The sampled protocol can invert the ranking of methods.}
On CA at $N=100$, SASRec scores HR@10 $93.24$ and CaST-POI $92.91$, so the sampled protocol would report SASRec as the better model, whereas under full ranking the same two runs are $44.25$ and $46.29$, a $2.04$-point advantage for CaST-POI that is significant at $p<0.01$ after correction. The same reversal appears on TKY at $N=100$ ($91.32$ against $91.27$). A pool of 100 drawn uniformly from ten thousand POIs almost never contains a near-miss, so it measures whether a model can reject obviously irrelevant locations, a question on which every method here is close to saturated. This is our reason for reporting full ranking throughout, and a caution about comparing across papers that use the sampled protocol.

\textbf{Scaling behaviour.}
On TKY and CA the advantage over the strongest baseline is absent or negative at small pools and clearly positive at full ranking: the HR@10 gap runs $-0.05$, $+1.14$, $+2.10$ and $+1.74$ on TKY at $N=100$, $1{,}000$, $2{,}000$ and full ranking, and $-0.33$, $-1.09$, $-0.89$ and $+2.04$ on CA. The trend is in the direction candidate conditioning predicts, namely that the more candidates are individually plausible, the more it matters that the user representation can be re-read for each of them, but it is not monotone and we do not claim it is. NYC behaves differently: against CORE the gap is $+0.89$ at $N=100$ and $+0.25$ at full ranking, shrinking rather than growing, consistent with the two methods not being distinguishable on that dataset.

\subsection{Robustness Analysis}

\textbf{Stratification.}
We stratify each test set along two axes. Users are split by the history visible at prediction time ($<50$, $50$--$100$, $100$--$200$, $\ge 200$ check-ins, giving $535$, $249$, $145$ and $115$ NYC instances), and target POIs by training-set visit frequency into \emph{tail}, \emph{torso} and \emph{head} at the $60$th and $80$th percentiles, where ties make the realised NYC buckets $640$, $198$ and $206$. Figure~\ref{fig:robustness} shows HR@10 per bucket on NYC for CaST-POI, CORE and SASRec.

\textbf{History length.}
On NYC, CaST-POI leads the best baseline of each bucket on the two shortest ($46.11$ against CORE's $45.36$ at ${<}50$, and $67.87$ against SASRec's $67.20$ at $50$--$100$) and trails on the two longest ($70.80$ against $75.40$ at $100$--$200$, and $75.94$ against $77.39$ at $\ge 200$). The model therefore helps where there is little history to compress and hurts where there is a lot, and the two effects roughly cancel, which is what the NYC null result consists of. On TKY and CA, where the aggregate gain is significant, CaST-POI leads every length bucket including $\ge 200$ ($70.75$ against SASRec's $69.03$ on TKY, $57.04$ against $56.19$ on CA).

\textbf{Target frequency.}
The gain sits on the head and torso, not the tail. On NYC, CaST-POI is $5.99$ points ahead of the best neural baseline on head ($61.65$ against SASRec's $55.66$) and $1.35$ ahead on torso ($70.37$ against $69.02$), but trails CORE on tail ($53.02$ against $55.73$). The pattern repeats on CA (head $62.74$ against $51.96$, tail $39.60$ against $39.98$) and softens on TKY, where CaST-POI leads all three buckets. Candidate-relative geometry helps most when a candidate has enough training signal for its embedding to be meaningful and the open question is which of several plausible places the user will go; for a POI with a handful of interactions there is little to condition on and popularity-driven baselines stay competitive. This is the opposite of the cold-start story such a method is often assumed to support.

The head bucket also exposes the benchmark itself. There the zero-parameter revisit heuristic reaches $64.08$ on NYC and $78.15$ on TKY, ahead of CaST-POI ($61.65$ and $73.73$) and of every neural baseline, because frequently-visited POIs are frequently \emph{re}-visited. The learned models earn their advantage on the torso and tail, where the heuristic collapses ($46.72$ on NYC tail, $25.35$ on TKY tail).

\section{Conclusion}
We presented \textbf{CaST-POI}, a POI ranker that grounds target attention in the geometry of human mobility through a candidate-relative spatial bias on the attention logits. Against seven sequential recommenders trained on an identical leave-one-out split with full-vocabulary ranking, it improves MRR by $7.6\%$ on TKY and $14.5\%$ on CA, both significant after multiple-comparison correction. The $5.1\%$ gain on NYC does not survive correction and we report it as null. Two findings qualify the design. The candidate-independent temporal bias has no measurable effect on any dataset and should be removed. The explicit revisit gate is the largest single component on NYC and TKY, which says more about the benchmarks than about our architecture, since next-POI as currently measured is substantially repetition prediction. Candidate-relative geometry helps, but only where it has room to discriminate, namely CA, head and torso POIs, and short histories. The sampled-candidate protocol, finally, reverses the ordering of methods on two of three datasets, which argues for full-vocabulary evaluation.

\bibliographystyle{ieeetr}
\bibliography{main2}

\end{document}